\documentclass[reprint,aps,prl]{revtex4-1}
\usepackage[utf8x]{inputenc}
\usepackage{graphics,graphicx,amssymb,amsmath,bm,dcolumn,url}% Include figure files% Align table columns on decimal point% bold math
\begin{document}
\title{Anomalous transport induced by nonhyperbolicity}
\author{S. R. Lopes$^{1}$}
\email{lopes@fisica.ufpr.br} 
\author{J. D. Szezech Jr$^2$}
\author{ R. F. Pereira$^3$}
\author{A. A. Bertolazzo$^{1,4}$}
\author{R. L. Viana$^1$}
\affiliation{$^1$Departamento de F\'{\i}sica, Universidade Federal do Paran\'a, Curitiba, PR, Brazil\\
$^2$Instituto de F\'{\i}sica, Universidade de S\~ao Paulo, S\~ao Paulo, SP, Brazil\\
$^3$Programa de P\'os-Gradua\c c\~ao em Ci\^encias/F\'{\i}sica, Universidade Estadual de Ponta Grossa, Ponta Grossa, PR, Brazil\\ 
$^4$Instituto de F\'{\i}sica, Universidade Federal do Rio Grande do Sul, Porto Alegre, RS, Brazil}

\date{\today} 

\begin{abstract}
In this letter we study how deterministic features presented by a system can be used to perform direct transport in a {\it quasi}-symmetric potential and weak dissipative system.  We show that the presence of nonhyperbolic regions around acceleration areas of the phase space plays an important role in the acceleration of particles giving rise to direct transport in the system. Such effect can be observed for a large interval of the weak asymmetric potential parameter allowing the possibility to obtain useful work from unbiased nonequilibrium fluctuation in real systems even in a presence of a {\it quasi}-symmetric potential.
\end{abstract}

                             % Classification Scheme.
\date{\today}
\maketitle

Anomalous transport is an emerging field in physics and, generally speaking, refers to nonequilibrium processes that cannot be described by using standard methods of statistical physics. The investigation of anomalous transport processes requires a combination of concept and methods of diverse disciplines, like stochastic theory, dynamical systems theory and disordered systems \cite{livro}. Anomalous transport occurs in a wide realm of physical systems ranging from a microscopic level (such as conducting electrons) to a macroscopic scale (as in global atmospheric events). One of the well-known phenomena in this category is anomalous diffusion, for which the mean-squared-displacement increases with time as a power-law $t^\mu$, where $\mu \ne 1$ \cite{feynman1996}. 
 
There is a growing interest in anomalous transport properties of nonlinear systems presenting nonequilibrium fluctuations, the {\it ratchet systems} \cite{mateos2000,wang2007,celestino2011,hutchings2004}. For these systems it is possible to surmount the second principle of thermodynamics provided we do not have any space or time symmetries forbidding direct transport \cite{astumian2002}. Ratchet systems occur in a variety of physical problems, like unidirectional transport in molecular motors \cite{astumian2002,veigel2009}, micro particles segregation in colloidal solutions \cite{rousselet1994}, and transport in quantum and nanoscale systems \cite{wang2007,linke2002,astumian2002}. 

Previous works have studied ratchet systems with a mixed phase space and weak dissipation. For such systems it has been shown that the ratchet effect results from a connection between weak dissipation, chaotic diffusion, and ballistic transport due to presence of periodic islands \cite{wang2007,celestino2011}. The latter, although occurring in the conservative case only, still affects the dynamics in the weak dissipative case since island centers become attracting fixed points in the weak dissipative case \cite{ulrike1996}. It has been shown that in the absent of islands anomalous transport cannot be achieved and the transport currents due to the ratchet effect in weakly dissipative systems are related to the existence of isoperiodic stable structures \cite{wang2007,celestino2011}. 

In these previous investigations it has been focused the strongly asymmetric case, for which the magnitude of potential asymmetry is comparable with the original potential energy of the system. In this letter we show that it is not necessary to have such strong asymmetry, in the sense that sizeable ratchet currents can be obtained in weakly dissipative systems with {\it slightly asymmetric} potentials. In fact, we claim that the presence of ratchet currents is influenced not so much by the potential asymmetry, but rather by the existence of {\it strongly nonhyperbolic regions} in the phase space of weakly dissipative systems. 

As a representative illustration of this effect we consider a periodically kicked rotor subjected to a harmonic potential function, whose dynamics is two-dimensional. By a hyperbolic region ${\cal S}$ we mean a set for which the tangent phase space in each point splits continuously into stable and an unstable manifolds which are invariant under the system dynamics: infinitesimal displacements in the stable (unstable) direction decay exponentially as time increases forward (backward) \cite{guckenheimer2002}. In addition, it is required that the angles between the stable and unstable directions are uniformly bounded away from zero. Chaotic orbits of dissipative two-dimensional mappings, for example, are often nonhyperbolic since the stable and unstable manifolds are tangent in infinitely many points. 

The dynamics of a periodically kicked rotor with small dissipation and potential asymmetry can be described in a cylindrical phase space $(-\infty\times\infty)\times [0,2\pi)$, whose discrete-time variables $p_n$ and $x_n$ are respectively the momentum and the angular position of the rotor just after the $n$th kick, with the dynamics given by the following dissipative asymmetric kicked rotor map (DAKRM) \cite{venegeroles2008}:  
\begin{eqnarray}
p_{n+1} &=& (1-\gamma) p_n - K[\sin(x_n) + a\sin(2x_n+\pi/2)], \\
x_{n+1} &=&  x_n + p_{n+1}, 
\end{eqnarray}
where $K$ is related to the kick strength, $0\le \gamma \le 1$ is a dissipation coefficient, and $a$ is the symmetry-breaking parameter of the system. The conservative ($\gamma = 0$) and symmetric ($a = 0$) limits yield the well-known Chirikov-Taylor map \cite{lichtenberg}. 
In the following we will keep the dissipation small enough (namely $\gamma =  2\times 10^{-4}$) in order to highlight the effect of the periodic islands of the conservative case. Moreover, the asymmetry parameter $a$ will be kept small so as to emphasize the role of the nonhyperbolic phase space regions on the anomalous transport.

The conservative and asymmetric ($a \ne 0$) case has been studied in detail: it has two fixed points (we call $P_1$) given by $x^{R,L} = \sin^{-1} \Theta(a,K)$ and $p^{R,L} = 0$, where $\Theta = \left(1-\sqrt{1+8a^2\pm 16\pi a/K}\right)/4a$, which are %marginally 
stable centers in the following parameter intervals: $0 < a <1/4$, and $6.40 < K < 7.20$. These two $P_1$ points are the centers of two %small
resonant islands that are actually accelerator modes. 
There are also two (left and right) period-$3$ fixed points ($P_3$) related to secondary resonances around the $P_1$ islands. 

In the weak dissipative case, the two $P_1$ points become stable foci, their basins of attraction being roughly the region occupied by the respective islands (in the conservative case). Moreover, the chaotic region in the conservative map becomes a chaotic transient in the weakly dissipative situation. At $K \approx 6.92$ the right $P_3$ points collides with the right fixed point $(p^{R}, x^{R})$ by a bifurcation. At the bifurcation point the small attraction basin of $P_1$ engulfs the stable manifold of the $P_3$ and turns to be accessible to points in a large phase space region. 

\begin{figure}
\includegraphics[width=\columnwidth]{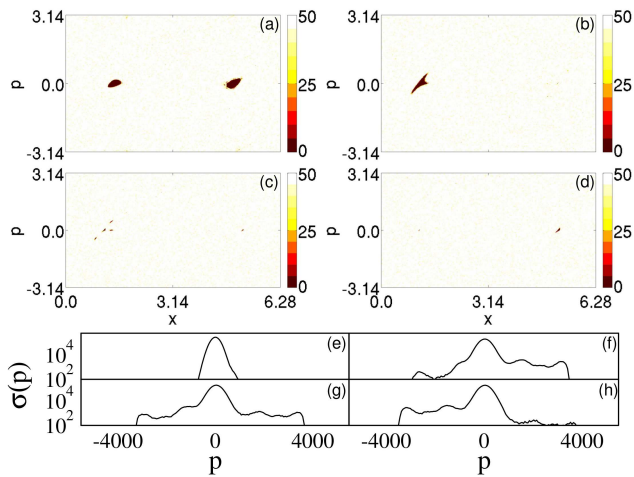}
\caption{\label{momento} (color online) 2-D histograms (top) and momentum probability distributions (bottom) for the DAKRM with $\gamma=0.0002$, $a=0.005$, and (a,e) $K=6.40$; (b,f) $6.92$; (c,g) $6.96$; and (d,h) $7.00$.}
\end{figure}

The vicinity of the fixed points plays a key role in the anomalous transport mechanism, in the same way as the islands do for the conservative case. More precisely, the wide accessibility of this vicinity near the bifurcation is responsible for large ratchet currents, just as the role of the accelerator modes in the non-dissipative map. Figs. \ref{momento}(a-d) depict 2-D histograms for $5000$ orbits (each orbit containing $10³$ points) of the DAKRM from initial conditions chosen in the phase plane region $0 < x < 2\pi$, $-\pi < p < \pi$, as well as the corresponding momentum probability distributions $\sigma(p)$ (Figs. \ref{momento}(e-h)). 
For $K = 6.40$ there is a quasi-symmetric situation, the neighborhood of the two $P_1$ fixed points (left and right) being seldom visited [Fig. \ref{momento}(a)].  
 Since the left (right) region is responsible for a negative (positive) increase of the momentum transport, we observe that for this parameter value the momentum distribution function is nearly symmetric, with a Gaussian shape [Fig. \ref{momento}(e)], resulting in a null transport.

Symmetry-breaking effects start to be noticeable after $K=6.40$ and turn to be maximum in the bifurcation at $K \approx 6.92$, where only the vicinity of the left $P_1$ is scarcely visited by orbits of the map [Fig. \ref{momento}(b)]. This effect is triggered by the bifurcation whereby the right $P_3$ fixed point collides with the right $P_1$ point and turns its vicinity easily accessible (not shielded), what is reflected in the asymmetric right tail in the momentum distribution function [Fig. \ref{momento}(f)]. The vicinity of the left  $P_1$ is not yet  affected since the collision process did not occur yet for  left $P_1$ and $P_3$. As a result, 
a net transport current is generated. However, this is more an effect of the bifurcation (due to nonhyperbolicity) than of the symmetry-breaking itself. In other words, if there is weak symmetry-breaking but no bifurcation (and no shield process), the ratchet effect will not occur, at least with the magnitude we observed in this example.

Not too far from the bifurcation, ($K = 6.96$), the vicinities of both $P_1$ fixed points become now regularly visited [Fig. \ref{momento}(c)] since positive and negative currents, caused by the right and left vicinity is counterbalanced. 
For this case  the momentum distribution function is again approximately symmetric [Fig. \ref{momento}(g)] but presenting right and left tails. The situation changes again after the left $P_1$ and $P_3$ fixed points collide [Fig. \ref{momento}(d)], through the same bifurcation mechanism described for their right counterparts. The increase in the accessibility of the vicinity of the left $P_1$ point leads to an asymmetric momentum distribution function [Fig. \ref{momento}(h)], restoring a net transport current. In this last case is noticeable that the seldom visited region is very small, nevertheless it is enough to inhibit positive currents. Once again the nonhyperbolicity of right region seems to be more important than the asymmetric situation itself.

\begin{figure}
\includegraphics[width=\columnwidth]{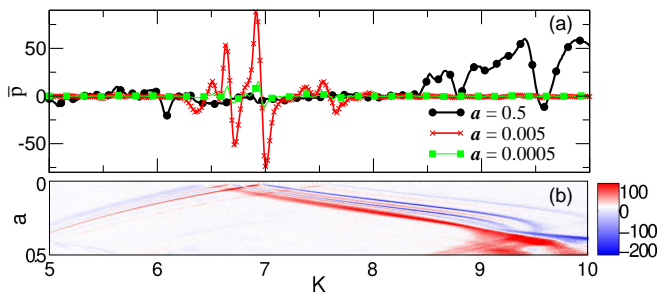}
\caption{\label{pmedio} (color online) (a) Ensemble averaged net current for different values of the nonlinearity and asymmetry parameters of the DAKRM with $\gamma = 0.0002$. (b) Net transport current (in colorscale) as a function of the asymmetry and nonlinearity parameters.}
\end{figure}

The variation of the average net transport current $\overline{p}$ with the nonlinearity parameter $K$ is depicted in Fig. \ref{pmedio}(a) for different values of the asymmetry parameter. The net current is ensemble-averaged over a large number ($10^6$) of initial conditions, each of them being followed by a short time $(t=10^3)$ to prevent the system to settle down into any fixed point. On varying $K$ we obtain a series of positive and negative net transport currents resulting from the ratchet effect. Curiously the net current fluctuates less for both very small and large asymmetry, being more sensitive to $K$ for intermediate values of $a$. As we have seen, the appearance of net currents is due to the fact that the left and right $P_3$ fixed points (that bifurcate in pairs for the symmetric case) start to bifurcate at {\it different} values of $K$. For the interval $6.40 < K < 7.00$, corresponding to Fig. \ref{momento}, the right $P_3$ fixed point bifurcates before the left one, generating a sequence of positive and negative net currents with very large peak values. For instance, the maximum amplitude of the net transport current for $a=0.005$ is at least three times larger than for similar parameters in the case of high asymmetry ($a=0.5$) \cite{wang2007}.

By the way of contrast, with a higher asymmetry value as those in Ref. \cite{wang2007} such large net currents are observed only for larger nonlinearities $(8.5<K<10)$, hence they are not primarily related to the bifurcations we present. In such case the role of the potential asymmetry overcomes the bifurcation mechanism presented here. Nevertheless appreciable net currents can also be acquired for asymmetry parameter as small as $a=0.0005$ or even smaller, although the peak values decrease considerably. Such decrease in the net transport current for small values of $a$ is expected since for the case of $a=0$ no transport can be observed due to the symmetry of the standard map. 

The sensitive dependence of the net transport current on the nonlinearity parameter in the weak asymmetry case reminds us of a similar behavior for the diffusion coefficient of the conservative and symmetric (Chirikov-Taylor) map, caused by the existence of accelerator modes. A more complete comparison is shown in Fig. \ref{pmedio}(b). If the potential asymmetry is large ($a \lesssim 0.5$) sizeable net currents are obtained only for higher values of $K$. This is due to the fact that large asymmetries shift the small islands present in the symmetric system phase space to some other intervals of $K$. On the other hand, if the asymmetry is very weak ($a \gtrsim 0$) we obtain large net currents (both positive and negative) for $K$-values considerably lower. The negative transport observed in Ref. \cite{wang2007} corresponds to the pale cyan area in Fig. \ref{pmedio}(b) for $a=0.5$ and around $K=6.5$, hence it is not directed correlated to the bifurcation mechanism we describe here. Moreover, the net current amplitude does not change appreciably along the blue and red lines in Fig. \ref{pmedio}(b) when $a$ is varied, emphasizing the role of the nonhyperbolic regions as an essential requirement for the ratchet effect to arise. 

The existence of nonhyperbolic regions in phase space, however small they may be, constitutes a {\it deterministic mechanism underlying anomalous transport} in the production of net currents through a ratchet effect. In order to quantify the degree of nonhyperbolicity related to the phenomena we describe in this letter, let us consider an initial condition $(p_0,x_0)$ and a unit vector ${\boldsymbol v}$, whose temporal evolution is given by ${\boldsymbol v}_{n+1}={\cal J}(p_n,x_n){\boldsymbol v}_n/|{\cal J}(p_n,x_n){\boldsymbol v}_n|$, where ${\cal J}(p_n,x_n)$ is the Jacobian matrix of the DAKRM. For $n$ large enough, ${\boldsymbol v}$ is parallel to the Lyapunov vector ${\boldsymbol u}(p,x)$ associated to the maximum Lyapunov exponent $\lambda_u$ of the map orbit beginning with $(p_0, x_0)$. Similarly a backward iteration of the same orbit gives us a new vector ${\boldsymbol v}_n$ that is parallel to the direction ${\boldsymbol s}(p,x)$, the Lyapunov vector associated to the minimum Lyapunov exponent $\lambda_s$ \cite{grebogi,ginelli}. 
For regions where $\lambda_s < 0 < \lambda_u$ the vectors ${\boldsymbol u}(p,x)$ and ${\boldsymbol s}(p,x)$ are tangent to the unstable and stable manifolds, respectively, of a point $(p,x)$. 

\begin{figure}
\includegraphics[width=\columnwidth]{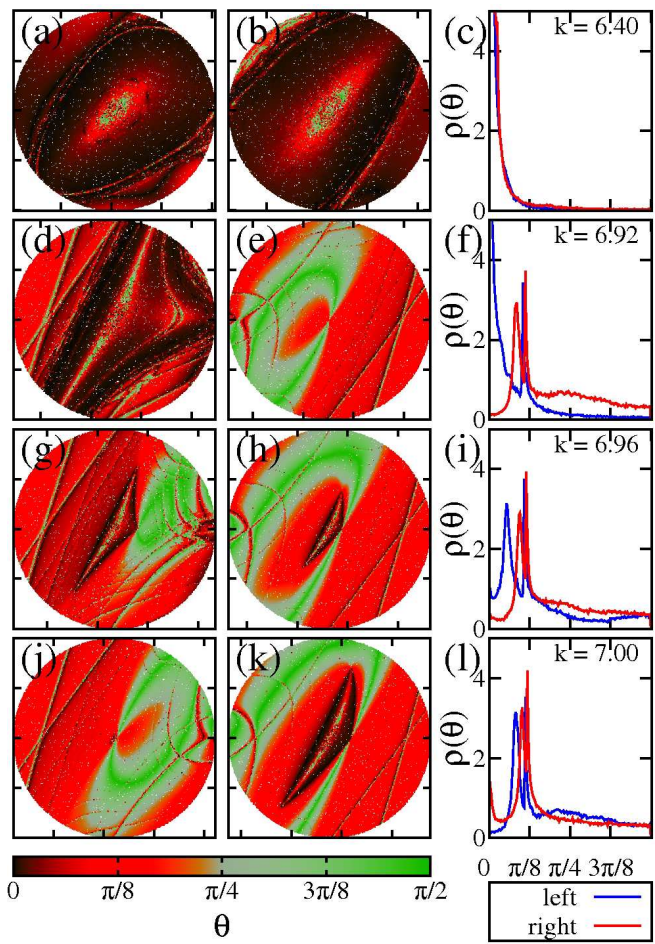}
\caption{\label{nonhyperbolic} (color online) $\theta(p,x)$ values  evaluated from $10^5$ initial conditions uniformly distributed around (0.2 radius) ${\cal S}_\epsilon^{R,L}$ and the distribution function of $\theta$.}
\end{figure}

The nonhyperbolic degree of a region ${\cal S}$ can be studied by computing the local angles between the two manifolds $\theta(p, x) = \cos^{-1} (| {\boldsymbol u}\cdot {\boldsymbol s}|)$, for $(p, x) \in {\cal S}$ \cite{ginelli}. So, $\theta(p,x) = 0$ denotes a tangency between unstable and stable manifolds at $(p,x)$.  
Let $S_\epsilon^{R,L} = \{ (p,x) \in \Omega : |(p-x) - (p^{R,L},x^{R,L}| < \epsilon \}$ be a $\epsilon$-radius neighborhood of the right and left $P_1$ fixed points. Results for $\theta(p,x)$ and its distribution function $\rho(\theta)$ calculated in both regions are shown in Fig. \ref{nonhyperbolic} for four values of the nonlinear parameter $K$. The dark regions in Fig. \ref{nonhyperbolic} correspond to strongly nonhyperbolic region surrounding the acceleration region.

For $K = 6.40$, near the fixed points $(p^{R,L},x^{R,L})$ there is a strong nonhyperbolic region which shields the acceleration area [Figs. \ref{nonhyperbolic}(a-b)]. In fact, almost all $\theta$ values for the right area (red curve) and left (blue curve) are confined in the interval $\theta < \pi/16$, configuring a strongly nonhyperbolic region around both fixed points [Fig. \ref{nonhyperbolic}(c)]. By way of contrast, when $K = 6.92$, only the left fixed point neighborhood is shielded, resulting in a large positive transport since only the right acceleration region is regularly visited [Figs. \ref{nonhyperbolic}(d-e)]. The right $P_1$ and $P_3$ fixed points suffer a bifurcation and all tangencies of stable and unstable manifolds disappear from the right area, allowing the trajectories to visit the acceleration region. Accordingly Fig. \ref{nonhyperbolic}(f) presents  different distributions of $\theta$ values for left and right regions. The red curve (right) presents a distribution peak  around $\theta=\pi/8$ considerably greater than the blue one, leading to absence of shielding in the right region.

In the case of $K = 6.96$ neither of the areas surrounding the fixed points are shielded by the tangencies of manifolds, resulting in large positive and negative transport currents, but no net current at all [Figs. \ref{nonhyperbolic}(g-i)]. This situation, however, is different from the one displayed by Figs. \ref{nonhyperbolic}(a-c), where the regions surrounding both fixed points were scarcely visited, hence there is no net current since the positive and negative currents are very small. 
Finally, for a higher value of $K$, only the right acceleration region is shielded, resulting in a negative transport current [Figs. \ref{nonhyperbolic}(j-k)]. The distribution $\rho(\theta)$ presents a peak  near zero for the red curve, confirming the existence of a shielded right area [Fig. \ref{nonhyperbolic}(l)]. 

The source of nonhyperbolicity in these regions is the tangencies between stable and unstable manifolds of saddle orbits embedded in the chaotic region therein. These tangencies prevent the system to visit the acceleration region. The scenario can be regarded as a counterpart of the Poincar\'e-Birkhoff' theorem (that describes the torus breakdown of a conservative two-degree of freedom map) for weakly dissipative systems.  Let the inner region surround the $P_3$ fixed point.  
When the stable and unstable manifolds of $P_3$ are almost tangent, the inner region surrounding $P_3$ mapped to an outer region tends to zero since the angle between the manifolds goes to zero. This clearly inhibits transport from inner to outer regions. As a result the acceleration region is practically not visited and no net transport currents are observed when there are nearly parallel manifolds. 

In conclusion we have shown that anomalous transport displayed by a {\it quasi}-symmetric potential and weakly dissipative system is strongly related to the topology of the acceleration regions around fixed points displayed by the system. The presence of nonhyperbolic regions caused by almost parallel unstable and stable manifolds can inhibit a chaotic trajectory to visit the neighborhood of the acceleration region surrounding fixed points of the system. This mechanism is closely related to the scenario described by the Poincar\'e-Birkhoff theorem in area-preserving two-dimensional maps. This dynamical phenomenon yields large net transport current in some direction even though the potential has an extremely small degree of symmetry-breaking. Hence such net currents can yield useful work from unbiased nonequilibrium fluctuation even with {\it quasi}-symmetric potentials, which enlarges the realm of dynamical systems displaying the ratchet effect.

This work is partially supported by CNPq, CAPES and Funda\c c\~ao Arauc\'aria.

\end{document}